# Spin Seebeck effect through antiferromagnetic NiO

Arati Prakash[1], Jack Brangham[1], Fengyuan Yang[1], Joseph P. Heremans[2,1,3]

[1]The Ohio State University Department of Physics, Columbus, Ohio 43210

[2]The Ohio State University Department of Mechanical Engineering, Columbus, Ohio 43210

[3]The Ohio State University Department of Materials Science and Engineering, Columbus, Ohio 43210

**Abstract**

We report temperature-dependent spin-Seebeck measurements on Pt/YIG bilayers and Pt/NiO/YIG trilayers, where YIG (Yttrium iron garnet, $Y_3Fe_5O_{12}$) is an insulating ferrimagnet and NiO is an antiferromagnet at low temperatures. The thickness of the NiO layer is varied from 0 to 10 nm. In the Pt/YIG bilayers, the temperature gradient applied to the YIG stimulates dynamic spin injection into the Pt, which generates an inverse spin Hall voltage in the Pt. The presence of a NiO layer dampens the spin injection exponentially with a decay length of 2±0.6 nm at 180 K. The decay length increases with temperature and shows a maximum of 5.5±0.8 nm at 360 K. The temperature dependence of the amplitude of the spin-Seebeck signal without NiO shows a broad maximum of 6.5±0.5 μV/K at 20 K. In the presence of NiO, the maximum shifts sharply to higher temperatures, likely correlated to the increase in decay length. This implies that NiO is most transparent to magnon propagation near the paramagnet-antiferromagnet transition. We do not see the enhancement in spin current driven into Pt reported in other papers when 1-2 nm NiO layers are sandwiched between Pt and YIG.

**Introduction**

In the spin-Seebeck effect (SSE, longitudinal configuration), [1] a heat current stimulates spin propagation across an interface between a ferromagnetic (FM) material and a normal metal (NM), where it is converted into a transverse electric field by the inverse spin Hall effect (ISHE).[2] The most commonly studied material system is the Pt/YIG system. Recently,[3,4,5] SSE signals have been reported in systems where the magnetic material is an antiferromagnet (AFM). In principle,[6] in the absence of a net spin polarization, an AFM cannot produce a spin-Seebeck signal. A net spin polarization can be produced by bringing the AFM above the spin-flop transition by an applied magnetic field, resulting in recent experimental reports of SSE signals in $Cr_2O_3$[3] and $MnF_2$,[4] as well as a theory for it.[7] The effect on $MnF_2$ is particularly large, reaching over 40 µV/K at high field and low temperature ($T$).[4] However, an uncertainty remains about the exact values of the temperature gradient applied to the AFM, and thus the absolute values of the spin-Seebeck coefficients. The effect on $Cr_2O_3$ was much smaller[3] and more in line with values hitherto observed on the Pt/YIG system.

We report here on SSE signals measured on a thin AFM layer grown on a FM, the Pt/NiO/YIG trilayer structure, which was recently reported to exhibit antiferromagnetic magnon transport.[8] In that study, microwave-driven ferromagnetic resonance (FMR) in the YIG FM layer was used to produce a pumped spin current that flowed through the NiO AFM layer and reached the Pt layer, where it produced an ISHE electric field. Monitoring the amplitude of this field as a function of NiO layers thickness (varied from 0 to 100 nm) gave information about the propagation of a dynamic net magnetic moment through the AFM. The results showed that magnonic transport through NiO layers thicker than 4 nm was attenuated exponentially with a decay length of about 10 ± 1.2 nm at room temperature. However, the insertion of a 2 nm thin NiO layer between YIG and Pt enhanced the spin currents driven into Pt. A subsequent theory[9] ascribes the origin of this effect to diffusion of thermal antiferromagnetic magnons in the NiO. If that is so, one would not expect to see the enhancement in the SSE measurements, and this is why the present experimental study was undertaken. We note that SSE results on the Pt/NiO/YIG system were recently published[5] after we started this study, in which an enhancement is reported. In contrast, we do not see the enhancement and we discuss the possible sources for that difference.

In previous studies,[10] the temperature dependence of the SSE in the YIG/Pt system without any NiO shows a broad maximum, attributed to the properties of the magnon dispersion. We report the same behavior here. This maximum was explained[10] by the fact that thermally driven magnons have a spectral distribution that covers almost the entire Brillouin zone at room temperature, unlike FMR-driven magnons at ~10 GHz near the zone center. The magnon dispersion[11] of YIG includes several optical-like branches that have practically no group velocity, and are not expected to contribute much to the SSE signal, but it also includes quasi-acoustic branches. The non-monotonic temperature dependence then arises as follows. At the temperatures below the maximum, the SSE signal is limited by the thermal magnon population; it increases with increasing temperature because higher energy magnons get excited, and in larger numbers. In this temperature range, most magnons have rather similar group velocity (only the magnons with energies near about 10 GHz have near-zero group velocity). At temperatures above the maximum, the effects of the magnon inelastic mean free path[12] and diffusion length[13] become more important, and the SSE now decreases with increasing temperature. It was further concluded from the magnetic-field dependence of the SSE measurements[10,14] at high field that the magnon modes at energies below 40 K contribute the most to the SSE effect in the Pt/YIG system. The present study includes the temperature dependence of the SSE in YIG/Pt, which shows the same behavior as in Ref. [10], and to this we add temperature-dependent data of the SSE through the AFM, which was not included in the FMR study.[8]

**Experiment**

YIG films were grown epitaxially on single crystal $Gd_3Ga_5O_{12}$ (GGG) <111> substrates by ultrahigh vacuum off-axis sputtering.[15,16,17] The YIG was grown at a substrate temperature of 750°C while being rotated at 10°/s for optimal sample uniformity at a growth rate of 0.51 nm/min. The thickness of YIG films grown on GGG substrates was 250 nm, as it has been shown[18] that the intensity of the SSE signal in YIG films grown on GGG increases with increasing YIG thickness, but saturates for thicknesses of about 250 nm. NiO and Pt layers were then deposited at room temperature on these YIG films also by off-axis sputtering. A series of Pt/NiO/YIG samples with length 10 mm and width 5 mm were deposited with 6 nm thick Pt and 0-10 nm thick NiO. In all cases, the Pt deposition was done *in-situ* with the NiO growth. A summary of the samples studied is given in Table I.

| Sample | $T_M$ (K) |
| --- | --- |
| Pt(6nm)/YIG(250nm) | 20 |
| Pt(6nm)/NiO(1nm)/YIG | 120 |
| Pt(6nm)/NiO(2nm)/YIG | 220 |
| Pt(6nm)//NiO(5nm)/YIG | 260 |
| Pt(6nm)//NiO(10nm)/YIG | 320 |

*Table I properties of the samples studied.*

The SSE measurements were conducted as in Ref. [10], except for the way the temperature gradient was calculated. High temperature measurements were conducted in a liquid nitrogen flow cryostat from 80 to 420 K. Samples were mounted using Apiezon H-grease ($T >$ 300 K) or N-grease ($T <$ 300 K) on electrically insulating cubic boron nitride (c-BN) heat-spreading pads with high in-plane thermal conductivity. Three resistive heaters (120 Ohm) were applied to the c-BN pads and connected electrically in series, ensuring uniform heating. To switch the SSE signal, we swept the field between ±50 mT. Low temperature measurements from 2 to 300 K were conducted on the Pt/YIG sample in a Physical Property Measurement System (PPMS) by Quantum Design. All experimental parameters were controlled as much as possible to be identical between the two systems. The voltage across the Pt layers was measured with a Keithley 2182 nanovoltmeter. Electric fields were calculated from raw traces (an example is shown in Fig. 1a for a Pt(6nm)/YIG(250nm)/GGG bilayer) of the voltage across the Pt strip, $V_y$, by averaging the values at -50 mT and +50 mT, then dividing by the length $L_y$ of the Pt layer to give the induced ISHE field ($E_y \equiv V_y/L_y$ in V/m).

Unlike in Ref [10], temperature gradients $\nabla T$ are calculated from the measured heater power output $Q$ per unit cross-sectional area of YIG (in units of W/ m$^2$) and the measured values for bulk thermal conductivity of the YIG[12] ($\kappa_{YIG}$) at each temperature: $\nabla T = Q/\kappa_{YIG}$. This procedure assumes that there is negligible loss of heat between the heater and the heat sink, so that all the heat goes though the sample as a uniform flux. Heat spreaders assured uniformity and an adiabatic sample mount minimized heat losses. The high thermal conductance of the sample, due to the high thermal conductivity of GGG and the favorable geometry, combined with the extensive use of radiation shielding and of 25 µm copper or manganin voltage and heater wires, ensure that heat losses are minimal. The procedure also assumes that the thermal conductivity of the 250 nm YIG film is the same as that of bulk YIG. In the absence of reliable data on the temperature dependence of the thermal conductivity of 250 nm-tick YIG films, this hypothesis is not as easily justified. The error may affect the absolute values reported for the spin-Seebeck coefficient $S_y$, but the relative effect of adding NiO layers between the YIG and the Pt is much less sensitive to this hypothesis, since the thickness of the YIG layer is held constant. The values of $S_y$ obtained following this procedure in different cryostats and in different runs are superimposed on each other in Fig. 1(b). The spread in the data, of the order of 30% at worst, gives a quantitative estimate of the accuracy with which we can make sample to sample comparisons, as we need to do to study the effect of NiO thickness. We developed this process, instead of using the method in Ref [10], because it is not sensitive to thermal contact resistances between films and substrate or between the different layers, while direct thermometry is. To test that, Quantum Design thermometry was mounted on the hot- and cold-side c-BN pads, but the resulting measurements were affected by thermal contact resistances between sample and heat sink, and proved to be less reproducible than the procedure used here. The values for the spin-Seebeck coefficient $S_y \equiv E_y / \nabla T$ (in units of nV/K), are reported as the ratios of electric field to temperature gradient, with the geometrical factors of the samples divided out. The procedure is consistent with our previous measurements,[10] enabling a quantitative comparison of all results.

**Results**

SSE signals are observed on all samples at room temperature. We begin with the case of the Pt/YIG sample (Table 1). We point out that the observed spin-Seebeck coefficients (Fig. 1b) in the Pt(6nm)/YIG(250 nm) bilayer on GGG are one to two orders of magnitude larger than our own previously reported signals for Pt on YIG single crystal,[10] but an important reason is simply the difference in the method used to calculate $\nabla T$. The maximum SSE signal is 6.5±0.5 μV/K near 20 K. These results demonstrate the critical role of the Pt/YIG interface: the pristine interface between Pt and epitaxial YIG film allows the conduction electrons in the Pt directly interact with the localized YIG magnetization in the absence of interfacial defects, which is highly desired for spin transfer across the interface. In contrast, for our previous[10] YIG bulk crystal with a polished surface (even for epi-ready YIG single crystals), it is expected that there is defect layer at the YIG surface. Given that our previous FMR spin pumping study[19] shows that a nonmagnetic insulating barrier of only 1 nm thickness at the interface can reduce the ISHE signal by a factor of ~250, it is reasonable to expect that any defect layer will strongly affect the results.

Figure 2 shows the SSE of the Pt/NiO/YIG trilayer systems as a function of temperature. The SSE signal is hard to discern from the noise floor of 50 nV at some low-temperature cutoff point, varying from 100 to 180 K depending on sample thickness: Fig. 2 reports only data above this cut-off temperature for each sample. The temperature dependence of SSE shows a maximum at a temperature $T_M$ that decreases monotonically with the NiO thickness, and is summarized in Table 1. $T_M$ is not necessarily the Néel temperature $T_N$ of NiO films, because two competing mechanisms are at work. On the one hand, it is known that $T_N$ in free NiO films is lower than that of bulk NiO and has some systematic variation with film thickness.[20] On the other hand, it was shown[21] by neutron diffraction on $CoO/Fe_3O_4$ films in the same thickness range as those studied here that the ordering temperature of the CoO films is enhanced for small CoO thicknesses. There is clearly an interaction between the Ni spins in NiO and the Fe spins in YIG by an exchange coupling or biasing effect. The positive slope of $dS_y/dT$ below $T_M$ indicates that the propagation of magnons though the NiO film improves as the thermal fluctuation in the AFM-ordered NiO is increased with increasing temperature. The negative slope of $dS_y/dT$ above $T_M$ in the Pt/NiO/YIG trilayers simply mirrors the negative slope behavior of the Pt/YIG system in the absence of NiO.

Figure 3a shows how the SSE signal decays exponentially (i.e. $S_y = S_0 \exp(-th/L)$ where $th$ is the NiO thickness, $S_0$ is a prefactor, and $L$ is an attenuation length) with increasing NiO thickness at 420, 360, 300, 240 and 180 K. We fit the attenuation length $L$ to the data and report it as a function of temperature in Fig. 3b. The decay continues to be observed down to 100 K, but the number of samples on which the SSE is measurable has decreased. This behavior and the attenuation length at 300 K is within about 40% of that observed[8] with FMR driven spin flux. At first sight this is surprising, if we consider that the modes excited in FMR-driven spin flux are near the Brillouin zone center and have a much larger spatial extent than the high energy modes exited in thermal spin fluxes. Recall that in FMR spin pumping of Pt/NiO/YIG trilayers, the excitation is uniform coherent precession of the YIG magnetization (or, zero-$k$ magnons), while for thermally-driven SSE measurements of the same structure, the excitation is thermal magnons that have finite $k$ and are diffusive. Therefore the ratio between the NiO film thickness and the magnon wavelength would be expected to be very different for FMR spin pumping and SSE; thus, the decay length would be expected to follow the same argument as well. The similar behaviors between the two experiments indicates that not all thermal magnons participate in the SSE, but only those at small $k$-values. This same conclusion was arrived at during the interpretation of the freeze-out of the SSE effect at high magnetic field.[10,14] The attenuation length decreases with decreasing temperature below 300 K, and seems to have a broad maximum near 360 K, although the decrease above 360 K is within experimental error bars. This hints that two separate mechanisms are at work here, one related to thermal fluctuations of the magnons, and one to the coupling between Ni spins in NiO and Fe spins in the underlying YIG.

Unlike what was observed with FMR-pumped magnons,[8] the SSE measurement of Pt/NiO/YIG for the 2 nm NiO reported here does not show an enhancement. This also contrasts with the recently reported SSE measurements on similar structures using polycrystalline YIG substrates,[5] where an enhancement is observed similar to our earlier report of FMR spin pumping. We believe these two SSE measurements do not contradict each other because of the difference in measurement techniques and sample characteristics. In this work we used 250-nm YIG epitaxial films and in Ref. 5, polished polycrystalline substrates were used. We also note that in Ref. 5, the maximum magnitude of $S_y$ for Pt(3 nm)/NiO(2 nm)/YIG(bulk) is about 1.1 µV/K at ~260 K, while in Fig. 2 of this work, the maximum $S_y$ for Pt(6 nm)/NiO(2 nm)/YIG(250 nm)/GGG is 0.35 µV/K at 200 – 300 K. Considering that the resistance of a 3 nm Pt layer is typically a few times higher than that of a 6 nm Pt layer, these results of the same measurements on two similar samples are consistent. We believe that the AFM magnons and spin fluctuations in NiO are likely responsible for the spin current enhancement through NiO in Ref. 5, similar to our earlier FMR spin pumping results.[8] The absence of enhancement in this work is likely due to the differences in YIG samples (epitaxial films vs. polycrystalline bulk) and interfacial characteristics, as well as fact that our Pt/YIG sample already exhibit a large SSE signal.

In summary, we report the transmission of thermally-driven magnons excited in the YIG films through a NiO layer in its antiferromagetic and its paramagnetic state, with an attenuation length ranging between 2 and 5.5 nm at temperatures from 180 to 420 K. From this we conclude that the thermally-driven magnons that give rise to the SSE must be limited to the low-energy part of the magnon spectrum, consistent with previous work on the high-field suppression of the SSE.

**Acknowledgement**

This work was primarily supported by the Army Research Office (ARO) MURI W911NF-14-1-0016 and the U.S. Department of Energy (DOE), Office of Science, Basic Energy Sciences, under Grants No. DE-SC0001304. It is partially supported by the Center for Emergent Materials, an NSF MRSEC under grant DMR-1420451.

**Figure captions**

Figure 1. (a) A raw trace of the spin-Seebeck voltage $V_y$ versus applied field at 300 K for an applied heater power of 0.495 W and (b) spin Seebeck Coefficient as a function of temperature for a Pt(6 nm)/YIG(250 nm) bilayer deposited on GGG <111>. In (b), the results of several different runs in different instruments are superimposed on each other, illustrating the reproducibility of the results, which is of the order of 30%. A maximum Spin Seebeck Coefficient of ~6.5±0.5 µV/K is observed on the YIG film within a broad peak around 20 K.

Figure 2. Temperature dependence of the SSE signal on the Pt/YIG bilayer is compared to SSE signals from the YIG/NiO/Pt trilayers, with the NiO film thickness as noted. A decrease in SSE signal is observed with increasing NiO film thickness. A maximum at a temperature $T_M$ is observed for each of the YIG/NiO/Pt trilayers. $T_M$ increases with increasing NiO thickness.

Figure 3. (a) Magnitude of the SSE signal at 420, 360, 300, 240, and 180 K plotted as a function of NiO thickness. The experimental values are the data points. The lines through them are fitted exponential decay functions, with a characteristic attenuation length scale given as a function of temperature in (b).

**Figure 1**

Fig. 1 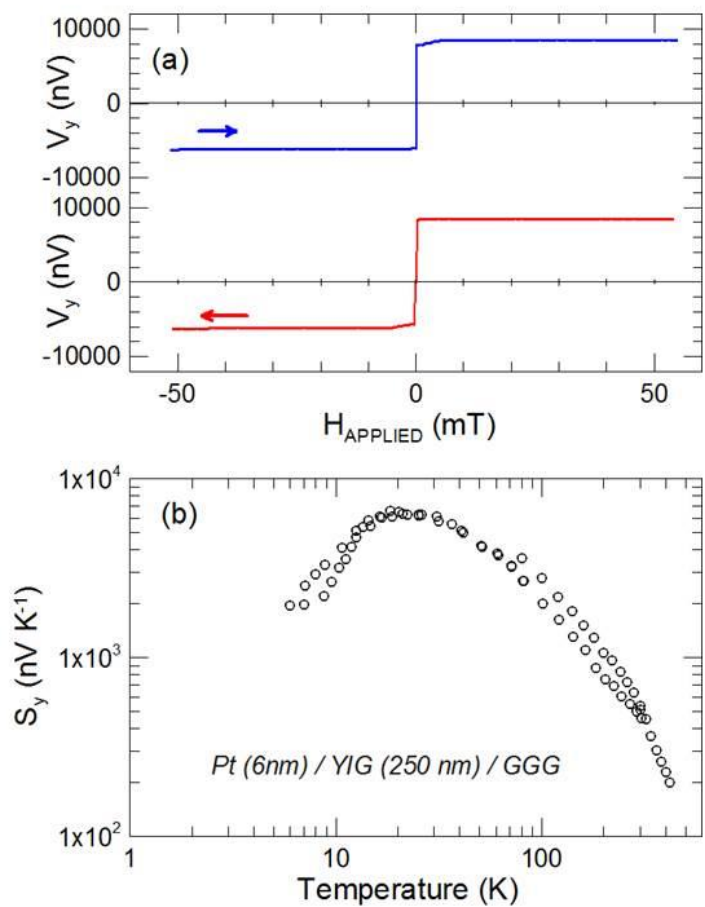

**Figure 2**

Fig. 2

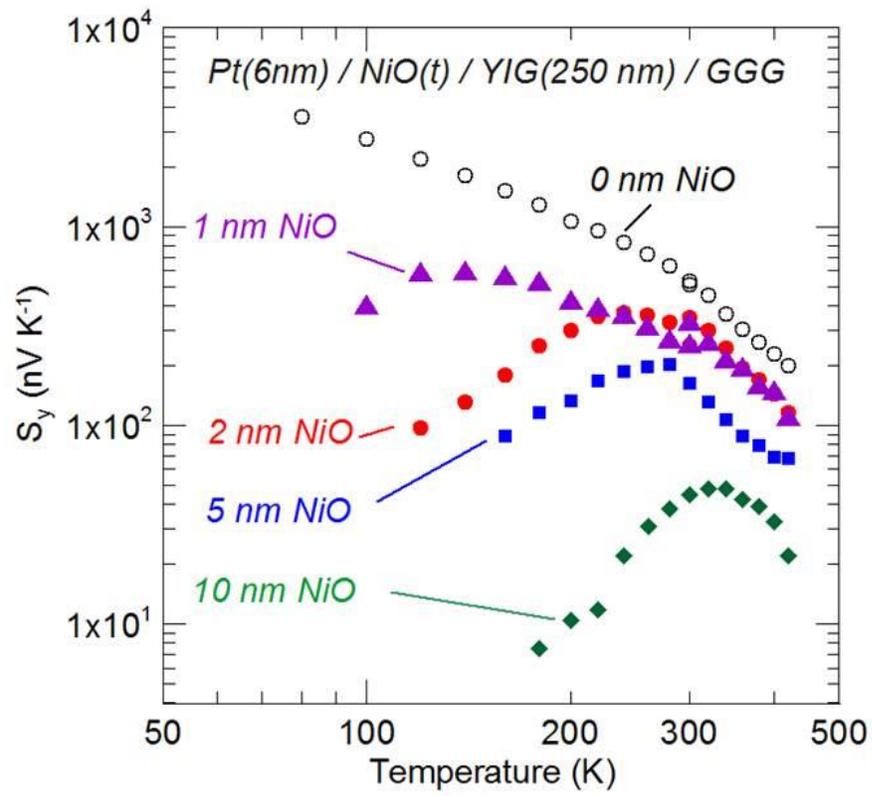

**Figure 3**

Fig. 3

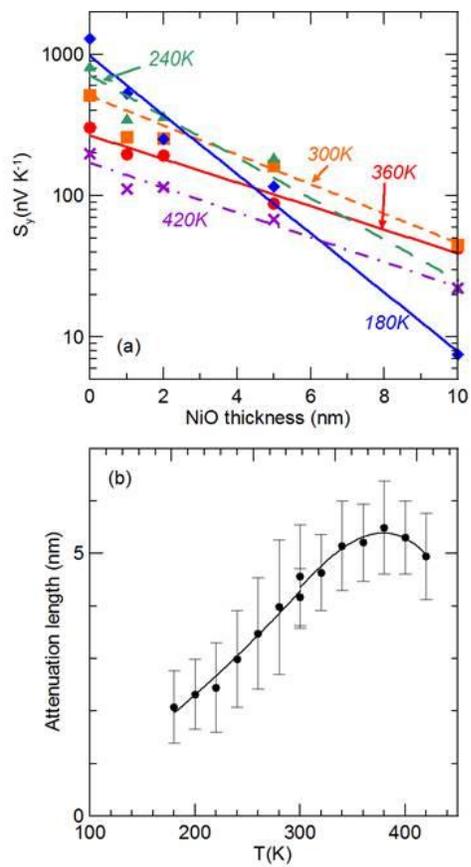